\documentclass[letterpaper,12pt]{article}
\usepackage{tabularx} 
\usepackage{amsmath}
\usepackage{amsmath,bm}

\usepackage{graphicx} 
\usepackage[margin=1in,letterpaper]{geometry} 
\usepackage{cite} 
\usepackage[final]{hyperref} 
\hypersetup{
	colorlinks=true,       
	linkcolor=blue,        
	citecolor=blue,        
	filecolor=magenta,     
	urlcolor=blue         
}
\usepackage{bbold}

\begin{document}

\title{Transforming nondepolarizing Mueller matrices into Jones matrices}
\author{M. A. Kuntman
and E. Kuntman}



\maketitle

\begin{abstract}

It is well known that there exists a four dimensional complex vector associated with a nondepolarizing Mueller matrix. In this note it is shown that this complex vector, which is isomorphic to the Jones matrix, can be obtained from the nondepolarizing Mueller matrix apart from an overall phase.
\end{abstract}


\section{Introduction}

In polarization optics Jones matrices are very useful for representing optical properties of deterministic optical media. When a Jones matrix, $\mathbf{J}$, is given it is transformed into a Mueller
matrix by the relation\cite{Simon,Kim}:

\begin{equation}
\mathbf{M}=\mathbf{A}(\mathbf{J}\otimes \mathbf{J^*})\mathbf{A^{-1}}
\end{equation}
where the superscript $^*$ indicates complex conjugation, the superscript $^{\dagger}$ indicates the Hermitian conjugate, $\otimes$ is the Kronecker product and the unitary matrices $\mathbf{A}$, $\mathbf{A^{-1}}$ are defined as follows: 

{\footnotesize
\begin{equation}
\mathbf{A}=\begin{pmatrix}
1&0&0&1\\
1&0&0&-1\\
0&1&1&0\\
0&i&-i&0\end{pmatrix},
\quad\mathbf{A}^{-1}=\frac{1}{2}\mathbf{A}^{\dagger}=\frac{1}{2}\begin{pmatrix}
1&1&0&0\\
0&0&1&-i\\
0&0&1&i\\
1&-1&0&0
\end{pmatrix},
\end{equation}}
Since the overall phase of the Jones matrix is lost during the transformation infinitely many
Jones matrices can be mapped into one Mueller matrix.

It is also desirable to be able to transform from a given Mueller matrix
into a Jones matrix. But, Jones matrices cannot represent depolarization. Thus, only non-depolarizing Mueller
matrices have corresponding Jones matrices (Mueller-Jones matrices). On the other hand, Mueller matrices cannot represent overall phase, therefore a one to one transformation is not possible from non-depolarizing Mueller matrices into  Jones matrices.
 
A method for transforming non-depolarizing Mueller matrices into Jones matrices was given by Chipman \cite{Chipman}. In this note it is shown that transformation of non-depolarizing Mueller matrices into Jones matrices can also be accomplished by means of a four dimensional complex vector which is isomorphic to the Jones matrix \cite{KKA,KKPA, KKCA}. 
\section{Vector state corresponding to Jones and Mueller-Jones matrices}

To study the properties of a Mueller matrix $\mathbf{M}$ (non-depolarizing or depolarizing) it is useful to transform $\mathbf{M}$ into a Hermitian matrix $\mathbf{H}$:

\begin{equation}\label{coh}
\mathbf{H}= \frac{1}{4}\sum_{i,j=0}^3M_{ij}(\bm{\sigma}_{i}\otimes\bm{\sigma}_{j}^*),
\end{equation}
where $M_{ij}$ are the elements of $\mathbf{M}$, $4\times4$ matrices $\bm{\sigma}_{i}\otimes\bm{\sigma}_{j}^*$ form a complete basis for a given $\mathbf{M}$ to $\mathbf{H}$ transformation and 
 $\bm{\sigma_{i}}$ are the Pauli matrices with the $2\times2$ identity matrix in the following order:
\begin{equation}
\bm{\sigma}_0=\begin{pmatrix}
1&0\\
0&1
\end{pmatrix},\quad\bm{\sigma}_1=\begin{pmatrix}
1&0\\
0&-1
\end{pmatrix},\\\bm{\sigma}_2=\begin{pmatrix}
0&1\\
1&0
\end{pmatrix},\quad\bm{\sigma}_3=\begin{pmatrix}
0&-i\\
i&0
\end{pmatrix}.
\end{equation}

The representation of $\mathbf{H}$ given by Eq. \eqref{coh} is not unique.  Instead of the basis $\bm{\sigma}_{i}\otimes\bm{\sigma}_{j}^*$ we can use the transformed basis \mbox{$\mathbf{U}(\bm{\sigma}_{i}\otimes\bm{\sigma}_{j}^*)\mathbf{U}^{-1}$}, where $\mathbf{U}$ is any suitable unitary transformation.
In this work we will let $\mathbf{U}= \mathbf{A}$ and we will use the following transformed basis:
\begin{equation}\label{prefer}
\bm{\Sigma}_{ij}=\mathbf{A}(\bm{\sigma}_{i}\otimes\bm{\sigma}_{j}^*)\mathbf{A^{-1}},
\end{equation}

In the new basis:
\begin{equation}\label{covb}
\mathbf{H}= \frac{1}{4}\sum_{i,j=0}^3M_{ij}\bm{\Sigma}_{ij},
\end{equation}
with an explicit matrix form:
{\footnotesize
\begin{equation}\label{Ha}
\mathbf{H}=\frac{1}{4}\left(\begin{array}{c|c|c|c}
M_{00}+M_{11}&M_{01}+M_{10}&M_{02}+M_{20}&M_{03}+M_{30}\\
M_{22}+M_{33}&-i(M_{23}-M_{32})&+i(M_{13}-M_{31})&-i(M_{12}-M_{21})\\
\hline
M_{01}+M_{10}&M_{00}+M_{11}&M_{12}+M_{21}&M_{13}+M_{31}\\
+i(M_{23}-M_{32})&-M_{22}-M_{33}&+i(M_{03}-M_{30})&-i(M_{02}-M_{20})\\
\hline
M_{02}+M_{20}&M_{12}+M_{21}&M_{00}-M_{11}&M_{23}+M_{32}\\
-i(M_{13}-M_{31})&-i(M_{03}-M_{30})&+M_{22}-M_{33}&+i(M_{01}-M_{10})\\
\hline
M_{03}+M_{30}&M_{13}+M_{31}&M_{23}+M_{32}&M_{00}-M_{11}\\
+i(M_{12}-M_{21})&+i(M_{02}-M_{20})&-i(M_{01}-M_{10})&-M_{22}+M_{33}\\
\end{array}\right)
\end{equation}}

If and only if the Mueller matrix of the system is non-depolarizing, the associated $\mathbf{H}$ matrix will be of rank 1. In this case it is always possible to define a  vector $|h\rangle$ such that
\begin{equation}
\mathbf{H}=| h\rangle\langle h|,
\end{equation}
where $|h\rangle$ is the eigenvector of $\mathbf{H}$ corresponding to the single non-zero eigenvalue \cite{Gil1, Gil2, Aelio, chipman2}.

In the basis $\bm{\Sigma}_{ij}$ dimensionless components of $|h\rangle$ can be parametrized as $\tau$, $\alpha$, $\beta$ and $\gamma$:
\begin{equation}
|h\rangle=\begin{pmatrix}
 \tau\\\alpha\\\beta\\\gamma
\end{pmatrix}.
\end{equation}
The parameters $\tau, \alpha$, $\beta$ and $\gamma$ are, in general  complex numbers, but one of them can always be chosen as real and positive if the overall phase is disregarded. 

Similarly, any Jones matrix can be also written in terms of $\tau, \alpha, \beta$ and $\gamma$:
\begin{equation}
\mathbf{J}= \tau\bm{\sigma}_0+\alpha\bm{\sigma}_1+\beta\bm{\sigma}_2+\gamma\bm{\sigma}_3 .
\end{equation}
so that
\begin{equation}
\mathbf{J}=
\begin{pmatrix}
\tau+\alpha & \beta-i\gamma\\
\beta+i\gamma& \tau-\alpha
\end{pmatrix}.
\end{equation}

As it was shown previously \cite{KKCA}, $|h\rangle$ vector and Jones matrix are isomorphic to each other; therefore, finding the components of $|h\rangle$ is enough to determine the Jones matrix.

\section{Transforming Mueller-Jones matrices into Jones matrices}

In terms of the parameters $\tau$, $\alpha$, $\beta$ and $\gamma$,  $\mathbf{H}$ matrix takes its simplest form:
\begin{equation}\label{Hb}
\mathbf{H}= |h\rangle\langle h| =
\begin{pmatrix}
\tau\tau^*&\tau\alpha^*&\tau\beta^*&\tau\gamma^*\\
\alpha\tau^*&\alpha\alpha^*&\alpha\beta^*&\alpha\gamma^*\\
\beta\tau^*&\beta\alpha^*&\beta\beta^*&\beta\gamma^*\\
\gamma\tau^*&\gamma\alpha^*&\gamma\beta^*&\gamma\gamma^*
\end{pmatrix}.
\end{equation}

$\mathbf{H}$ matrix has four column vectors with the following properties:

\begin{equation}
|c_1\rangle=\tau^*|h_1\rangle,\quad |c_2\rangle=\alpha^*|h_2\rangle,\quad |c_3\rangle=\beta^*|h_3\rangle,\quad |c_4\rangle=\gamma^*|h_4\rangle.
\end{equation}

$\mathbf{H}$ corresponds to a non-depolarizing Mueller matrix  ($rank(\mathbf{H})=1$), hence,  any of the column vectors, $|c_i\rangle$, can be chosen as the linearly independent one. On the other hand, $\langle h_1|h_1\rangle=\langle h_2|h_2\rangle=\langle h_3|h_3\rangle=\langle h_4|h_4\rangle$, therefore, $|h_i\rangle$ differ from each other only by respective phases.

The column vectors $|c_i\rangle$ can be read from Eq.\eqref{Ha} in terms of the elements of the non-depolarizing Mueller matrix:

\begin{equation*}
|c_1\rangle=\frac{1}{4}\begin{pmatrix}M_{00}+M_{11}+M_{22}+M_{33}\\M_{01}+M_{10}+iM_{23}-iM_{32}\\M_{02}+M_{20}-iM_{13}+iM_{31}\\M_{03}+M_{30}+iM_{12}-iM_{21}
\end{pmatrix},\quad |c_2\rangle=\frac{1}{4}\begin{pmatrix}M_{01}+M_{10}-iM_{23}+iM_{32}\\M_{00}+M_{11}-M_{22}-M_{33}\\M_{12}+M_{21}-iM_{03}+iM_{30}\\M_{13}+M_{31}+iM_{02}-iM_{20}
\end{pmatrix},\qquad     
\end{equation*}

\begin{equation}
|c_3\rangle=\frac{1}{4}\begin{pmatrix}M_{02}+M_{20}+iM_{13}-iM_{31}\\M_{12}+M_{21}+iM_{03}-iM_{30}\\M_{00}-M_{11}+M_{22}-M_{33}\\M_{23}+M_{32}-iM_{01}+iM_{10}
\end{pmatrix},\quad |c_4\rangle=\frac{1}{4}\begin{pmatrix}M_{03}+M_{30}-iM_{12}+iM_{21}\\M_{13}+M_{31}-iM_{02}+iM_{20}\\M_{23}+M_{32}+iM_{01}-iM_{10}\\M_{00}-M_{11}-M_{22}+M_{33}
\end{pmatrix}.     
\end{equation}

As an example, suppose that an unnormalized non-depolarizing Mueller matrix is given as follows:

\begin{equation}
\mathbf{M}=\begin{pmatrix}
20&-2&-10&6\\
10&-8&-10&14\\
2&14&-2&8\\
-6&-2&16&6
\end{pmatrix}
\end{equation}

From the transformation given in Eq.\eqref{Ha} the associated $\mathbf{H}$ matrix can be found:

\begin{equation}
\mathbf{H}=\begin{pmatrix}
4&2+2i&-2+4i&6i\\
2-2i&2&1+3i&3+3i\\
-2-4i&1-3i&5&6-3i\\
-6i&3-3i&6+3i&9
\end{pmatrix}
\end{equation}

Any nonzero column vector of $\mathbf{H}$ can be used in order to find the components of $|h\rangle$. In this example all column vectors of $\mathbf{H}$ are nonzero, therefore it is possible to use any column vector to find the components of $|h\rangle$. For example, the first column vector, $|c_1\rangle$, can be written as:

\begin{equation}
|c_1\rangle=\tau^*|h_1\rangle=\begin{pmatrix} 4\\2-2i\\-2-4i\\-6i
\end{pmatrix}
\end{equation}

Choosing $\tau$ to be real and positive, i.e., taking $\tau=2$, unnormalized $|h_1\rangle$ becomes:

\begin{equation}
|h_1\rangle=\begin{pmatrix}
2\\1-i\\-1-2i\\-3i
\end{pmatrix},
\end{equation}
where $\tau=2, \: \alpha=1-i,\: \beta=-1-2i, \: \gamma=-3i$ and $|h_1\rangle$ can be normalized by the factor $\sqrt{\langle h_1|h_1\rangle}=\sqrt{20}$.

If $\tau$ happens to be zero, any other nonzero column vector, $|c_i\rangle$ can be used to find the components of $|h_i\rangle$. For example, if $\alpha\neq 0$:

\begin{equation}
|c_2\rangle=\alpha^*|h_2\rangle=\begin{pmatrix}
2+2i\\2\\1-3i\\3-3i
\end{pmatrix}
\end{equation}

Choosing $\alpha$ to be real and positive, i.e., taking $\alpha=\sqrt{2}$, unnormalized $|h_2\rangle$ becomes:

\begin{equation}
|h_2\rangle=\begin{pmatrix}
(2+2i)/\sqrt{2}\\2/\sqrt{2}\\(1-3i)/\sqrt{2}\\(3-3i)/\sqrt{2}
\end{pmatrix}
\end{equation}
Similarly, unnormalized $|h_2\rangle$ vector can be normalized by the same factor $\sqrt{\langle h_2|h_2\rangle}=\sqrt{\langle h_1|h_1\rangle}=\sqrt{20}$. 

In this case the components of the unnormalized $|h_2\rangle$ are $\tau=(2+2i)/\sqrt{2}$,\: $\alpha=2/\sqrt{2}$,\: $\beta=(1-3i)/\sqrt{2}$, \: $\gamma=(3-3i)/\sqrt{2}$, and they look different than the parameters found for $|h_1\rangle$. However, this is not an unexpected result:  $|h_1\rangle, |h_2\rangle, |h_3\rangle, |h_4\rangle$ are not identical to each other; they are equivalent to each other and they differ from each other only by respective phases. In the present example:

\begin{equation}
|h_1\rangle=\frac{1-i}{\sqrt{2}} |h_2\rangle.
\end{equation}

\section{Conclusion
}
In polarization optics Jones matrices are very useful for representing optical properties of a non-depolarizing medium. When a Jones matrix is given it can be transformed into a Mueller
matrix. It is also desirable to be able to transform from a given Mueller matrix
into a Jones matrix. But, Jones matrices cannot represent depolarization. Thus, only non-depolarizing Mueller
matrices have corresponding Jones matrices (Mueller-Jones matrices).
A method for transforming non-depolarizing Mueller matrices into Jones matrices was already given by Chipman. In this note it is shown that transformation of non-depolarizing Mueller matrices into Jones matrices can also be accomplished by means of a four dimensional complex vector which is isomorphic to the Jones matrix.


\end{document}